\documentclass[%
 reprint,
 amsmath,amssymb,
 aps,
]{revtex4-1}

\usepackage{graphicx}
\usepackage{dcolumn}
\usepackage{bm}

\begin{document}

\preprint{APS/123-QED}

\title{Spin entanglement and nonlocality of
multifermion systems
}

\author{Arifullin Marsel}
 \altaffiliation[]{arifullinm@mail.ru}
\author{Berdinskiy Vitaly}%
 \email{bvl@unpk.osu.ru}
\affiliation{%
 Physics Department, Orenburg University, Orenburg, Pobedy avenue, 13, Russia
 }%

\date{\today}

\begin{abstract}
Spin density matrices of the system, containing arbitrary even number N of indistinguishable fermions with spin S = 1/2, described by antisymmetric wave function, have been calculated. The indistinguishability and the Pauli principles are proved to determine uniquely spin states, spin correlations and entanglement of fermion spin states. Increase of the particle number in the multifermion system reduces the spin correlation in any pair of fermions. The fully entangled system of N electrons are shown to be composed by pairs with nonentangled spin states that is the incoherent superposition of the singlet and triplet states. Any large system of N fermions, such as electrons with spin S = 1/2, the spin state of any particle are shown to be entangled with the other part of the system containing N-1 particle. However, the spin state of this electron is not entangled with any other particle, and spin state of any electron pair is not entangled. These properties of spin states manifest in the Einstein-Podolsky-Rosen as confirmation or violation of the Bell inequalities indicating the presence of non-local quantum spin correlations.
\begin{description}

\item[PACS numbers] 03.65.Ud, 05.30.Ch, 75.25.-j

\end{description}
\end{abstract}

\pacs{Valid PACS appear here}
\maketitle


\section{\label{sec:level1}Introduction}

The fundamental property of fermions – the spin S =1/2 – as well as the charge determines both individual and collective properties, for example, the symmetry of multifermion wave functions and statistical properties of ensembles. The indistinguishability principle applied for bosons leads to the Pauli's principle and the main property of multifermion wavefunctions – antisymmetry. Any antisymmetric wave function is known can be presented as the sum of  Slater's determinants of one fermion functions. The problem of entanglement in multifermion systems attracted a lot of attention during last decades and was the main aim of many investigations \cite{Schli01,Eckert,Amico}.
Possible applications of electron spin as the information carrier in spintronics \cite{Fabian}, quantum computing and quantum cryptography require the knowledge of spin states of multifermion systems such as electrons in semiconductors, superconductors, spin liquids, etc. \cite{Valiev,Kitaev,Cirac95}. However, for using of the electron spin as the quantum information carrier, it should be extracted from an ensemble of indistinguishable particles. If the extraction process is fast enough, then the electron spin has no time to change its spin state, and therefore saves the memory about its presence in the large ensemble. Thus, the knowledge of multispin states and their properties, such as spin correlations and spin entanglement is needed for above mentioned applications.
The entanglement is the important characteristics of quantum states, that is needed for algorithms of quantum calculations and protocols of quantum cryptography. These facts have determined, on one hand, active theoretical investigations of entanglement, and, on the other, underestimation of the importance of entanglement for descriptions of real physical systems and processes. The physical meaning of quantum state entanglement is followed from the main property of the density matrix of  entangled systems \cite {Kilin}
\begin{equation}
\rho ^{AB} \ne \sum _{i}p_{i} \rho _{i}^{A} \otimes \rho _{i}^{B}
\end{equation}
here \textit{$\rho^{A} $} and \textit{$\rho^{B} $}  are  density matrices of  subsystems  \textit{A} and \textit{B}. This inequality   means that density matrix \textit{$\rho^{AB} $}  of united system can not be obtained as the sum of direct production of density matrices  \textit{$\rho^{A} $} and \textit{$\rho^{B} $} . Therefore, the full system cannot be created as simple unification of independent physical subsystems. However, if the complex system can be created by unification of independent subsystems, then it is nonentangled. Thus, entangled systems creation needs specific selection rules, that manage control physical processes leading to entangled complex systems. The example of such selection rules are spin selection rules that determine formation of singlet (entangled) particles from precursors having noncorrelated electron spins \cite{Zel}.

The classical examples of entangled states are Bell's biparticle states \cite{Bouw}. A lot of treatises have been devoted to investigation of entangled states, but most of them considered simple two spin models \cite{Wang}; the number of known three spin models is limited. Besides, the entanglement in multispin systems is approximately unknown yet in spite of the fact that multifermion entanglement can play the important role in condensed matter physics \cite{Lunk05}.

Information meaning of quantum state entanglement is known and described well in scientific literature \cite{Niel20,Val05}. However, the problem of entanglement genesis did not attract a lot of attention.  Moreover, this problem did not appear in most treatises where different kinds of entanglement were studied. As main sources of entanglement the Quolomb or exchange interactions were thought or implicated \cite{Ved03,Oh04}. Strictly speaking, the exchange interaction arises in the case of space overlapping of fermion wave functions, and manifestations of exchange interactions are followed from the Pauli's principle which requires the antisymmetry of fermion wave functions $\Psi $.
As the result of numerous theoretical investigation the common opinion insists that the antisymmetric wavefunctions which can be presented by Slater determinants rank 1 describe nonentangled (separate) state. For example, a pure state of two fermions presented as the Slater's determinant
\[{\left| \psi  \right\rangle} =2^{-1/2} \left\{{\left| \varphi _{1} \left(1\right) \right\rangle} {\left| \varphi _{2} \left(2\right) \right\rangle} -{\left| \varphi _{2} \left(1\right) \right\rangle} {\left| \varphi _{1} \left(2\right) \right\rangle} \right\},\]
(${\left| \varphi _{1} \left(i\right) \right\rangle} $ and ${\left| \varphi _{2} \left(i\right) \right\rangle} $ are orthogonal single particle states) is thought to be  nonentangled. However, if one considers one fermion wavefunctions as the production of space and spin parts, e.g. ${\left| \varphi _{i} \left(1\right) \right\rangle} =\phi \left(r_{i} \right){\left| s \right\rangle} $, the Slater's determinant  takes the well known form
\[\begin{array}{l} {{\left| \Psi \left(r,s\right) \right\rangle} =2^{-1/2} \det \left|\varphi _{1} (r,s)\varphi _{2} (r,s)\right|=} \\ {=2^{-1/2} \det \left|\begin{array}{cc} {\phi \left(r_{1} \right){\left| \alpha _{1}  \right\rangle} } & {\phi \left(r_{2} \right){\left| \alpha _{2}  \right\rangle} } \\ {\phi \left(r_{1} \right){\left| \beta _{1}  \right\rangle} } & {\phi \left(r_{2} \right){\left| \beta _{2}  \right\rangle} } \end{array}\right|=} \\ {=2^{-1/2} \left(\phi \left(r_{1} \right)\phi \left(r_{2} \right)\right)\left({\left| \alpha _{1}  \right\rangle} {\left| \beta _{2}  \right\rangle} -{\left| \beta _{1}  \right\rangle} {\left| \alpha _{2}  \right\rangle} \right)} \end{array}\]
here ${\left| \alpha _{i}  \right\rangle} $ and ${\left| \beta _{i}  \right\rangle} $ - the projection of the spin S = 1/2 on the axis OZ. The spin subsystem is evident to be in the singlet state ${\left| S_{12}  \right\rangle} =\left(2\right)^{-1/2} {\left| \alpha _{1} \beta _{2} -\beta _{1} \alpha _{2}  \right\rangle} $, which is classic example of the entangled Bell's state. This simple example proves that the entanglement properties of subsystems can differ than ones of a whole system, and  spin subsystems of  indistinguishable fermions require separate consideration, as their properties do not follow automatically from the properties of the complete system.
Aims of this work are to calculate fermion  multispin density matrices in  forms which allow generalization for all kinds of fermions having spin S=1/2, to study properties of these density matrices  and  described spin states including spin correlations and entanglement in multifermion systems.

\section{\label{sec:level2}Multifermion spin states}

The comprehensive description of multifermion systems requires generally the knowledge of wave function $\Psi $, which depends on all independent coordinates of the system.  However, to describe physical states and properties of the spin subsystem the spin density matrix is needed only \cite{Landau,Blum}. It is shown below that for calculation of the spin density matrix the main property of the wave function $\Psi $ -- antisymmetry -- is necessary only. Any antisymmetric wave function is known to be presented as the superposition of Slater's determinants composed of wave functions, which depend on space and spin variables of all \textit{N} indistinguishable particles. However, for the sake of simplicity we will suppose that the one determinant wave function is enough to describe the whole fermion system. For \textit{N} fermions with spin S =1/2 occupying N/2 lowest states (wave functions are $\psi _{1} (r,s)$,$\psi _{2} (r,s)$,\dots $\psi _{N/2} (r,s)$) the Slater's determinant is
\begin{equation}
    {\left| \Psi \left(r,s\right) \right\rangle} =\left(N!\right)^{-1/2} \det \left|\psi _{1} (r,s)\psi _{2} (r,s)...\psi _{N/2} (r,s)\right|,
\end{equation}
here $\psi _{i} (r,s)=\varphi _{i} (r_{j} ){\left| s_{j}  \right\rangle} $ ($\varphi _{i} (r)$- describes the space part, and ${\left| s_{j}  \right\rangle} $- the spin part of the wavefunction). Spin density matrix$\rho ^{N} $ describing spin properties can be calculated from  $\rho ={\left| \Psi \left(r_{j} ,s_{j} \right) \right\rangle} {\left\langle \Psi \left(r_{j} ,s_{j} \right) \right|} $  by taking trace over all space coordinates and space wavefunctions $\varphi _{i} (r)$ of the whole system
\begin{eqnarray}{} {\rho ^{N} =Tr_{\varphi (r)} {\left| \Psi \left(r_{j} ,s_{j} \right) \right\rangle} {\left\langle \Psi \left(r_{j} ,s_{j} \right) \right|} =} \nonumber\\ {=\sum _{k}\left\langle \Phi _{k} {\left| \Psi \left(r_{j} ,s_{j} \right) \right\rangle} {\left\langle \Psi \left(r_{j} ,s_{j} \right) \right|} \Phi _{k} \right\rangle  }, \end{eqnarray}
here ${\left\langle \Phi _{k} \left(r_{i} \right) \right|} $- are direct productions of space wave functions $\varphi _{i} (r)$ describing  all possible transpositions of fermions. Calculation of the trace  \textit{Tr$\varphi $}   assumes the orthogonality of space wavefunctions   $\varphi _{i} (r)$.

 After calculation of the Slater's determinant by the Laplase method and taking trace over space wavefunctions the reduced spin density matrix $\rho ^{N} $ can be presented as the sum of nonorthogonal projection operators onto multispin singlet states, and for the system of \textit{N} fermion spins is

\begin{equation} \label{4} \rho ^{N} =\frac{2^{N/2} (N/2)!}{N!} \sum _{P}P\left({\left| S_{ij} S_{kl} S_{mn} ... \right\rangle} {\left\langle S_{ij} S_{kl} S_{mn} ... \right|} \right)  \end{equation}

The sum in equation \eqref{4} includes all possible placements of \textit{N} fermions on \textit{N/2} two-particle singlet spin states ${\left| S_{kl}  \right\rangle} =2^{-1/2} {\left| \uparrow _{k} \downarrow _{l} -\downarrow _{k} \uparrow _{l}  \right\rangle} =2^{-1/2} {\left| \alpha _{k} \beta _{l} -\beta _{k} \alpha _{l}  \right\rangle} $. Operator \textit{P} means permutations of fermion spins on all pair singlet states. The number of these summands is $2^{-N/2} N!/(N/2)!$, and is equal to the number of the Rumer's fermion pairings \cite{Rumer}. The expression \eqref{4} presents the spin density matrix $\rho ^{N} $  of the multifermion system. It does not depend on concrete space wavefunctions $\varphi_{i}(r_{i})$. The density matrix $\rho ^{N} $ is evidently to be determined by the indistinguishability of quantum particle and the Pauli's principle.

The presentation of the spin density matrix  $\rho ^{N} $   as the sum of nonorthogonal projection operators makes evident it's symmetry under  any transpositions of  fermions or their spins. Transpositions of two spins inside separated singlet state change the sign of the spin vector ${\left| S_{ij}  \right\rangle} =2^{-{1/2} } {\left| \alpha _{i} \beta _{j} -\beta _{i} \alpha _{j}  \right\rangle} $, but does not change the sign of their tensor production ${\left| S_{ij}  \right\rangle} {\left\langle S_{ij}  \right|} $.  Transpositions of spins from one singlet state to another one are equivalent to transpositions of projection operators, and do not change the spin density matrix \eqref{4} as a whole. As far as any pair spin states are invariant under any rotations, then the whole spin density matrix $\rho ^{N} $  is invariant under rotations too. Thus, the symmetric spin density matrix $\rho ^{N} $ is shown can be calculated if the antisymmetric wavefunction $\Psi \left(r,s\right)$ is known.

\subsection{\label{sec:level2}Spin state of four fermion system}

The four fermion system is the simplest nontrivial system which can be used to illustrate main properties of more complex systems.  The spin density matrix $\rho ^{4} $  can be calculated directly from the wavefunction presented as the Slater's determinant without using the formula \eqref{4}, and after some transformations it takes the form:
 \begin{eqnarray}\label{5} {\rho ^{4} =3^{-1} ({\left| S_{12} S_{34}  \right\rangle} {\left\langle S_{12} S_{34}  \right|} +{\left| S_{13} S_{24}  \right\rangle} {\left\langle S_{13} S_{24}  \right|} +} \nonumber \\ { +{\left| S_{14} S_{23}  \right\rangle} {\left\langle S_{14} S_{23}  \right|} )}, \end{eqnarray}
here the density matrix operator $\rho ^{4} $ is presented as the sum of three projection operators on singlet spin states ${\left| S_{ij} S_{kl}  \right\rangle} $. For the four-spin system the scalar products of vectors ${\left| S_{ij} S_{kl}  \right\rangle} $ are:

\begin{eqnarray*}
{\left\langle S_{12} S_{34} |  S_{13} S_{24}  \right\rangle =2^{-1},
\left\langle S_{12} S_{34} |  S_{14} S_{23}  \right\rangle} =-2^{-1}, \nonumber \\ {
{\left\langle S_{13} S_{24} |  S_{14} S_{23}  \right\rangle} =2^{-1}}.
\end{eqnarray*}

As far as these vectors describe nonorthogonal spin states, the density matrix in formulae \eqref{5} is presented as the sum of 3 non-orthogonal projection operators

\begin{eqnarray*}
P_{1} ={\left| S_{12} S_{34}  \right\rangle} {\left\langle S_{12} S_{34}  \right|} , P_{2} ={\left| S_{13} S_{24}  \right\rangle} {\left\langle S_{13} S_{24}  \right|} , \nonumber \\ {
P_{3} ={\left| S_{14} S_{23}  \right\rangle} {\left\langle S_{14} S_{23}  \right|}}.
\end{eqnarray*}

Non-orthogonal spin vectors are linearly dependent; and any vector can be presented as the superposition of two others. This means that they belong to the two-dimensional subspace of the full 16-dimentional spin space (the dimension of the  space is 24 = 16) and can be presented as usual vectors on the plane.

Another form of the spin density matrix operator can be obtained by introducing the other set of vectors:

\begin{eqnarray*}
{\left| 1 \right\rangle} ={\left| S_{12} S_{34}  \right\rangle} , {\left| 2 \right\rangle} ={\left| S_{13} S_{24}  \right\rangle} , {\left| 3 \right\rangle} ={\left| S_{14} S_{23}  \right\rangle} , \nonumber \\ {
{\left| 4 \right\rangle} =3^{-1/2} \left({\left| S_{13} S_{24}  \right\rangle} +{\left| S_{14} S_{23}  \right\rangle} \right)}
\end{eqnarray*}

Direct calculation shows that the spin vector ${\left| 4 \right\rangle} $ is normalized and orthogonal to the vector ${\left| 1 \right\rangle} ={\left| S_{12} S_{34}  \right\rangle} $. This new set of vectors allows presentation of  the spin density matrix $\rho ^{4} $   in the orthogonal basis as:

\begin{equation} \label{6}
 \rho ^{4} =\frac{1}{2} ({\left| 1 \right\rangle} {\left\langle 1 \right|} +{\left| 4 \right\rangle} {\left\langle 4 \right|} ) \end{equation}

Equality \eqref{6} makes it evident that $\rho ^{4} $ is proportional to the two-dimensional  identity matrix \textit{I} in the subspace of four spin singlet states. It is convenient to present the vector ${\left| 4 \right\rangle} $ as

\[{\left| 4 \right\rangle} =3^{-1/2} {\left| T_{12}^{+} T_{34}^{-} +T_{12}^{-} T_{34}^{+} -T_{12}^{0} T_{34}^{0}  \right\rangle} \]
here ${\left| T_{\pm ,0}  \right\rangle} $ - vectors of the pair triplet states

\[{\left| T_{ij}^{+}  \right\rangle} ={\left| \alpha _{i} \alpha _{j}  \right\rangle} ,  {\left| T_{ij}^{0}  \right\rangle} =2^{-1/2} {\left| \alpha _{i} \beta _{j} +\beta _{i} \alpha _{j}  \right\rangle} , {\left| T_{ij}^{-}  \right\rangle} ={\left| \beta _{i} \beta _{j}  \right\rangle} .\]

The density matrix \eqref{6} describes the simple noncoherent superposition of two  four-spin states, whose total spins are \textit{S} = 0. Expression \eqref{6} allows to calculate easily the value of the von Neumann entropy $S=-Tr\left(\rho \ln \rho \right)$, which is used often for estimations of the entanglement. Simple calculations shows that $S_4 = \ln 2$.

Multiplication of the equation \eqref{6} by 2 gives the two-dimensional identity operator \textit{I} in the right part. This operator is also the projection operator onto the two-dimensional singlet subspace. As far as equalities \eqref{5} and \eqref{6} describe the same operator, so the projection operator $P_{0}$ on the two-dimensional subspace can be represented as the sum of non-orthogonal projection operators
\begin{eqnarray}
{P_{0} =\frac{2}{3} ({\left| S_{12} S_{34}  \right\rangle} {\left\langle S_{12} S_{34}  \right|} +{\left| S_{13} S_{24}  \right\rangle} {\left\langle S_{13} S_{24}  \right|} +} \nonumber \\ { +{\left| S_{14} S_{23}  \right\rangle} {\left\langle S_{14} S_{23}  \right|} )},
\end{eqnarray}
This expression could be very useful for analysis of spin effects in multispin systems, because it can describe easily multispin selection rules that operate inevitably in many physical and chemical processes and reactions. This presentation of the projection operator can be generalized easily to more complex cases of multispin singlet states.

\subsection{\label{sec:level2}The spin density matrix of two fermion subsystems}

Spin systems of two fermions, for example, two electrons are most studied objects in the theory of quantum entanglement. Theoretical investigations of such two-spin models had been appeared very heuristic for producing new ideas and conceptions. However, the number of the ``pure'' two-spin systems is not so much: the helium atom, the hydrogen molecule, and the deuteron nuclei. In all other cases any two-fermion system, for example, the two electron system, should either be extracted from large system or should be considered as the subsystem of many-electron system. In both cases it is necessary to know, firstly, the spin states of the real two-electron systems, extracted from the large "mother system'', and, secondly, to know their difference from properties of the "ideal" and the well-studied two-spin system.

Below we consider the two-spin system, which was initially part of large ensemble of indistinguishable fermions and then was extracted from this ensemble. The ensemble is supposed to be in the ground state and is described by the density matrix \eqref{4}. According to the basic concepts of quantum mechanics, to describe all properties of the two-spin subsystem it will be enough to know the reduced two-spin density matrix $\rho^{12}$. This matrix can be calculated as the trace of $\rho^{N}$ over spin variables of all "extra" particles.

\begin{equation} \label{8}
\rho ^{12} =Tr_{N-2} \left(\rho ^{N} \right)
\end{equation}

As far as all the particles are indistinguishable and equivalent all particles having numbers N $>$2 will be considered as extra ones. To calculate $\rho^{12}$ the spin density matrix $\rho^{N}$  should be presented as the sum of two polynomials: the first one includes only the terms with operators ${\left| S_{12}  \right\rangle} {\left\langle S_{12}  \right|} $ (two spins belong to the same singlet state), and the second -- only terms with operators ${\left| S_{1k} S_{2l}  \right\rangle} {\left\langle S_{1k} S_{2l}  \right|} $ where spins $S_{1}$ and $S_{2}$ belong to different singlet pairs. The number of summands in the first polynomial can be easily determined by usual combinatorial rules, and their number is

\[\frac{(N-2)!}{2^{(N-2)/2} ((N-2)/2)!} .\]

After calculation of the trace all summands of the first polynomial give the following term in desired density matrix $\rho $12

\begin{equation} \label{9} \left(N-1\right)^{-1} {\left| S_{12}  \right\rangle} {\left\langle S_{12}  \right|}  \end{equation}

The numbers of summands in the second polynomial can be found in similar way, and the result is

\[\frac{(N)!}{2^{N/2} (N)/2)!} -\frac{(N-2)!}{2^{(N-2)/2} ((N-2)/2)!} \]

Calculation of the trace for summands of the second polynomial allows to find other terms of density matrix $\rho^{12}$

\begin{eqnarray} \label{10}
{4^{-1} \left(N-2\right)\left(N-1\right)^{-1} ({\left| S_{12}  \right\rangle} {\left\langle S_{12}  \right|} +}\nonumber \\ {+{\left| T_{12}^{+}  \right\rangle} {\left\langle T_{12}^{+}  \right|} +{\left| T_{12}^{0}  \right\rangle} {\left\langle T_{12}^{0}  \right|} +{\left| T_{12}^{-}  \right\rangle} {\left\langle T_{12}^{-}  \right|} )=} \nonumber\\ {=4^{-1} \left(N-2\right)\left(N-1\right)^{-1} I_{1} \otimes I_{2} }
\end{eqnarray}

Combining formulas \eqref{9} and \eqref{10} one can obtain  finally the density matrix $\rho ^{12} $ that describes the incoherent superposition of the singlet and triplet states of two fermion system included in or extracted from the N-femion system.

\begin{eqnarray} \label{11}
{\rho ^{12} =4^{-1} \frac{(N+2)}{\left(N-1\right)} {\left| S_{12}  \right\rangle} {\left\langle S_{12}  \right|} +4^{-1} \frac{(N-2)}{\left(N-1\right)} }\nonumber \\ {\times \left({\left| T_{12}^{+}  \right\rangle} {\left\langle T_{12}^{+}  \right|} +{\left| T_{12}^{0}  \right\rangle} {\left\langle T_{12}^{0}  \right|} +{\left| T_{12}^{-}  \right\rangle} {\left\langle T_{12}^{-}  \right|} \right)} \end{eqnarray}

The ratio of singlet and triplet states is dependent on the total number of fermions \textit{N}. The only system of two electrons (N = 2) can be in the pure singlet state, and described by the density matrix $\rho ={\left| S_{12}  \right\rangle} {\left\langle S_{12}  \right|} $. In all other cases (even N $>$ 2) any subsystem of two indistinguishable fermions will be in the spin state, which is noncoherent superposition of the singlet and triplet states.

The spin density matrix \eqref{9} can be used for calculations of correlation coefficients \textit{r} for two spins \textit{S} = 1/2 in ensembles of any even numbers of fermions.
\[\left\langle \left. \vec{S}_{1} \vec{S}_{2} \right\rangle \right. =Tr\left(\vec{S}_{1} \vec{S}_{2} \rho _{12} \right)/\left(\left|\vec{S}_{1} \right|\cdot \left|\vec{S}_{2} \right|\right)=-\left(N-1\right)^{-1} ,\]
here $\left|\vec{S}_{i} \right|=\sqrt{S(S+1)} =3^{1/2} /2$. The sign ``minus'' means, that the probability to find antiparallel orientations of fermion spins is always larger than the probability of the parallel orientation. Generally, the correlation coefficient \textit{r} depends on the number of spins in ensembles only, it is maximal for two fermions ($N = 2$), and is minimal if $N\rightarrow \infty$.

\[\mathop{\lim }\limits_{N\to \infty } \left\langle \left. \vec{S}_{1} \vec{S}_{2} \right\rangle \right. =-\mathop{\lim }\limits_{N\to \infty } \left(N-1\right)^{-1} =0\]

In the infinitely large system ($N\rightarrow \infty$) the spin state of the two-fermion subsystem is described by the density matrix

\begin{eqnarray} \label{12}
{\rho ^{12} {\rm (N}\to \infty {\rm )}=4^{-1} ({\left| S_{12}  \right\rangle} {\left\langle S_{12}  \right|} +{\left| T_{12}^{+}  \right\rangle} {\left\langle T_{12}^{+}  \right|} +} \nonumber\\ {+{\left| T_{12}^{0}  \right\rangle} {\left\langle T_{12}^{0}  \right|} +{\left| T_{12}^{-}  \right\rangle} {\left\langle T_{12}^{-}  \right|} )=4^{-1} I_{1} \otimes I_{2} }
\end{eqnarray}

This state is evident to be the noncoherent superposition of the spin states of two independent non-polarized fermions. Consequently, the increase of particle number in the multifermion system reduces correlation between spins of any fermion pair, and these correlations are absent if $N\rightarrow \infty$.

At the end of this section it is useful to note that the state of two-spin system is determined by the four-subsystem density matrix
\begin{eqnarray}\label{13}
{\rho ^{12} (N=4)=\frac{1}{2} {\left| S_{12}  \right\rangle} {\left\langle S_{12}  \right|} +}\nonumber \\
{+\frac{1}{6} \left({\left| T_{12}^{+}  \right\rangle} {\left\langle T_{12}^{+}  \right|} +{\left| T_{12}^{0}  \right\rangle} {\left\langle T_{12}^{0}  \right|} +{\left| T_{12}^{-}  \right\rangle} {\left\langle T_{12}^{-}  \right|} \right)}
\end{eqnarray}
This spin density matrix is known \cite{Aldos} to describe the unentangled state as far as it can be presented as the sum of direct productions of single-spin density matrices.

\section{\label{sec:level1}The entanglement of multispin fermion states}

Multifermion spin systems, which are described by the operator of the spin density matrix \eqref{4}, can be separated into two or more subsystems. Subsystems can have arbitrary dimensions, but their total dimension should be equal to the dimension of the initial system.  As examples of such separation can be mention different spontaneous decays of atomic nuclei, the processes of photoionization, transfer of electrons from the valence band into the conductivity one, etc.  So the question arises are spin systems of reaction products entangled or not? For example, for the semiconductor spintronics it is important to know are spin states of conductivity electrons entangled and are their spin states entangled with spin states of electrons which are left in the valence band? Can such entanglement of spin states  be determined by common genesis from valence band electrons?

The convenient criterion of entanglement is the Peres-Horodecki criterion \cite{Peres,Horod96}, which establishes connection between entanglement of subsystems A and B and presence of negative eigenvalues $\lambda_{i}$ for partially transposed density matrices $\rho ^{T_{B} } (AB)$. According to this criterion, for two subsystems A and B be entangled, it is necessary and sufficient that the partially transposed matrix $\rho ^{T_{B} } (AB)$ should has, at least, one negative eigenvalue $\lambda_{i}$. However, the presence of negative eigenvalues is equivalent to the statement that $\rho ^{T_{B} } $  is no longer the density matrix which should be nonnegative.  Therefore, for entangled states the partial transposition operation of the density matrix can not correspond to any real physical process.

The Peres-Horodecki criterion has appeared to be very convenient for the analysis of simple systems. For example, for the four-spin system, described by the spin density matrix \eqref{5},  the partially transposed density matrix has few negative eigenvalues among all possible   $\lambda_{i} = (1/2, 1/6, 1/6, 1/6, 1 / 6, 1/6, 1/6, -1 / 6, -1 / 6, -1 / 6)$. So, the entanglement measure \textit{E}, determined in accordance with \cite{Vidal02,Niel99} as the doubled sum of negative eigenvalues $\lambda_{i}$, is
\begin{equation*}
 E=-2\sum _{i}(\lambda _{i} ) =-2(-1/6-1/6-1/6)=1\,
\end{equation*}
This result means that for the four-spin system the entanglement between two-spin subsystems is maximal one, similar to the entanglement between two spins in Bell's singlet state.

However, the Peres-Horodecki criterion is hardly applied for investigation of large systems described by density matrices of higher dimensions \cite{Belous}, as far as analytical calculations of eigenvalues are impossible. Therefore, for studying of entanglement in large systems, similar to spin systems of multifermion systems, another criterions are needed. The existence of negative eigenvalues for partly transposed matrices $\rho ^{T_{B} } (AB)$  is equivalent to violation of the matrix nonegativity condition: for entangled systems the matrix $\rho ^{T_{B} } (AB)$ is not the positively defined matrix. Therefore, to prove the existence of entanglement between large multispin subsystems it is sufficient to prove violation of the positivity of the matrix $\rho ^{T_{B} } (AB)$ . It can be done by using, for example, the Sylvester criterion \cite{Gant}. Among the different definitions of the Sylvester criterion the most efficient is the requirement of non-negativity of all principal minors of the matrix, for example, the principal minors of the second order
\begin{equation} \label{14} M=\rho _{ii}^{T} \rho _{jj}^{T} -\rho _{ij}^{T} \rho _{ji}^{T} \, =\, \, \rho _{ii}^{T} \rho _{jj}^{T} \, -\, \left|\rho _{ij}^{T} \right|^{2} >0 \end{equation}
To prove violation of the Sylvester criterion for matrix $\rho ^{T_{B} } $ it is convenient to present the original density matrix $\rho (AB)$  \eqref{4} in the multiplicative basis as the block matrix $\rho (S_{z} ,S_{z}^{'} )$, where $S_{z} $ and $S_{z}^{'} $ - are projections of all possible multiplicative spin states of the complete system. Obviously, the only non-zero block of  such "extended" density matrix $\rho (S_{z} ,S_{z}^{'} )$  is the block corresponding to $S_{z} =0$ and $S_{z}^{'} =0$. For spin states with $S_{z} \ne 0$ and $S_{z}^{'} \ne 0$  all matrix elements (diagonal $\rho _{ii} $ and non-diagonal $\rho _{ij} $ ones)  in other blocks are equal to zero.

Multiplicative basis is set of orthogonal basis vectors ${\left| i \right\rangle} $ and ${\left| j \right\rangle} $, each of them is the direct product of  individual spin vectors ${\left| \uparrow  \right\rangle} $ or ${\left| \downarrow  \right\rangle} $. Simultaneously, any vectors ${\left| i \right\rangle} $ and ${\left| j \right\rangle} $  can be presented as multiplicative spin vectors of the subsystems $A$ and $B$
\begin{eqnarray*}
{\left| i \right\rangle} =\, {\left| 1 \right\rangle} \otimes {\left| 2 \right\rangle} \otimes \dots \otimes {\left| N \right\rangle}  = {\left| m_{A}  \right\rangle} \otimes {\left| l_{B}  \right\rangle}  \\
{\left| j \right\rangle} =\, {\left| 1' \right\rangle} \otimes {\left| 2' \right\rangle} \otimes \dots \otimes{\left| N' \right\rangle} \, \, =\, \, {\left| n_{A}  \right\rangle} \otimes {\left| k_{B}  \right\rangle}
\end{eqnarray*}
here ${\left| m_{A}  \right\rangle} $and $\, {\left| n_{A}  \right\rangle}$ are multiplicative spin vectors of the subsystem A, and multiplicative spin vectors ${\left| l_{B}  \right\rangle}$ and ${\left| k_{B}  \right\rangle}$ characterize  the subsystem $B$.  Both subsystems are of arbitrary dimensions $N_{A}$ and  $N_{B}$ , but $N_{A}$ + $N_{B}$ = $N$. The sum of spin projections for subsystems $A$ and $B$ satisfy the condition $S_{z}^{A} +S_{z}^{B} =0$ for all pairs of multiplicative vector (${\left| m_{A}  \right\rangle} $,${\left| l_{B}  \right\rangle} $) and (${\left| n_{A}  \right\rangle} $,${\left| k_{B}  \right\rangle} $) as far as vectors \textit{${\left| l_{B}  \right\rangle} $} and \textit{ ${\left| k_{B}  \right\rangle} $}  belong to the multispin singlet subspace. However, states ${\left| m_{A}  \right\rangle} $ and ${\left| n_{A}  \right\rangle}$ of the same subsystem A, and states ${\left| l_{B}  \right\rangle} $ and ${\left| k_{B}  \right\rangle}$ of the subsystem B have generally different sets of individual spin vectors , and their spin projections are not always equal to zero.  Moreover, vectors with $S_{z}^{} (m_{A} )\ne 0$, $S_{z}^{} (l_{B} )\ne 0$, $S_{z}^{} (n_{A} )\ne 0$ and $S_{z}^{} (k_{B} )\ne 0$ are always presented in the full set of spin vectors of subsystems $A$ and $B$.

If $S_{z}^{} (l_{B} )\ne S_{z}^{} (k_{B} )$and both $S_{z}^{} (l_{B} ),S_{z}^{} (k_{B} )\ne 0$, then the partial transposition ${\left| l_{B}  \right\rangle} \leftrightarrow {\left| k_{B}  \right\rangle} $ changes the spin state and spin projection   $S_{z}^{} $ as far as
\[S_{z}^{} (m_{A} )+S_{z} (l_{B} )\ne S_{z} (m_{A} )+S_{z} (k_{B} ),\]
and
\[S_{z}^{} (n_{A} )+S_{z} (k_{B} )\ne S_{z} (n_{A} )+S_{z} (l_{B} ).\]
This fact can be easily illustrated by the six spin system (N = 6).
For example,
\[{\left| i \right\rangle} = {\left| m_{A}  \right\rangle} \otimes {\left| l_{B}  \right\rangle} = {\left| \uparrow \uparrow \uparrow  \right\rangle} _{A} \otimes {\left| \downarrow \downarrow \downarrow  \right\rangle} _{B} \]
and
\[{\left| j \right\rangle} = {\left| n_{A}  \right\rangle} \otimes {\left| k_{B}  \right\rangle} = {\left| \downarrow \downarrow \downarrow  \right\rangle} _{A} \otimes {\left| \uparrow \uparrow \uparrow  \right\rangle} _{B} \]
Partial transposition of spin vectors ${\left| l_{B}  \right\rangle} \Leftrightarrow {\left| k_{B}  \right\rangle} $ changes whole vectors into new ones
\[{\left| i \right\rangle} \Rightarrow {\left| i' \right\rangle} =\, \, \, {\left| m_{A}  \right\rangle} \otimes {\left| k_{B}  \right\rangle} =\, {\left| \uparrow \uparrow \uparrow  \right\rangle} _{A} \otimes {\left| \uparrow \uparrow \uparrow  \right\rangle} _{B} \]
and
\[{\left| j \right\rangle} \Rightarrow {\left| j' \right\rangle} =\, \, \, {\left| n_{A}  \right\rangle} \otimes {\left| l_{B}  \right\rangle} =\, {\left| \downarrow \downarrow \downarrow  \right\rangle} _{A} \otimes {\left| \downarrow \downarrow \downarrow  \right\rangle} _{B} .\]
The total spin for both new states is $S = 3$ and spin projections are $S_{z}^{} =\pm 3$. As the result, the partial transposition
 ${\left| l_{B}  \right\rangle} \Leftrightarrow {\left| k_{B}  \right\rangle} $
 transfers none-zero off-diagonal matrix elements
 \begin{equation*}
  \rho _{ij} {\left| i \right\rangle} {\left\langle j \right|} = \rho _{ij} {\left| m_{A} l_{B}  \right\rangle} {\left\langle n_{A} k_{B}  \right|}
 \end{equation*}
from block $(S_{z} = 0$ and  $S'_{z} = 0)$ into off-diagonal matrix elements
\begin{equation*}
\rho _{i'j'}^{T} {\left| i' \right\rangle} {\left\langle j' \right|}  = \rho _{ij} {\left| m_{A} k_{B}  \right\rangle} {\left\langle n_{A} l_{B}  \right|}
\end{equation*}
in block $(S_{z} \ne 0$, $S'_{z} \ne 0)$. However, the partial transposition ${\left| l_{B}  \right\rangle} \Leftrightarrow {\left| k_{B}  \right\rangle} $ does not change diagonal elements
\begin{equation*}
\rho _{i'i'}^{T} {\left| i' \right\rangle} {\left\langle i' \right|} = \rho _{ii} {\left| m_{A} k_{B}  \right\rangle} {\left\langle m_{A} k_{B}  \right|} =\rho _{ii} {\left| i \right\rangle} {\left\langle i \right|}
\end{equation*}
neither in $(S_{z} = 0$ and  $S'_{z} = 0)$ block, nor in ($(S_{z} \ne 0$, $S'_{z} \ne 0)$ block, where all diagonal elements $\rho _{ii}^{T}$ , $\rho _{ii}^{T} $ are equal to zero before and after the partial transposition. Thus, the partially transposed matrix $\rho ^{T_{B} } (AB)$ has negative principal minors
\begin{equation*}
M=\rho _{ii}^{T} \rho _{ii}^{T} -\rho _{ij}^{T} \rho _{ji}^{T} =-\left|\rho _{ji}^{T} \right|^{2} <0.
\end{equation*}
So, it is not the positively defined matrix, and has, at least, one negative eigenvalue $\lambda $. This fact proves that in accordance with the Peres-Horodecki criterion the initial spin density matrix $\rho ^{N} $ \eqref{4} describes the entangled spin states of indistinguishable fermions, and the entanglement exists between all spin subsystems.

\section{\label{sec:level1}Manifestations of quantum correlations of multispin entangled states}

Physical properties of multispin states of the indistinguishable fermions, such as electrons, described by the density matrix $\rho^{N}$, allow to predict results of experiments (hypothetical, at least) which are interest both for the general theory of entanglement, and for quantum informatics. The thought $EPR$ experiment of  Einstein, Podolsky and Rosen \cite{epr} is the example of such experiments.  The theoretical analysis of this and similar experiments had been often used for investigations of fundamental problems of quantum mechanics, and, in particular, for verifying the Bell's inequalities \cite{Claus}. These inequalities describe correlations of some normalized physical parameters $Q$ and $R$ obtained by the observer $A$, and parameters $S$ and $T$ obtained  by the another observer B during studying decay of some physical system.  In general, the Bell inequality are written as follows
\begin{equation} \label{15} \left\langle QS\right\rangle +\left\langle RS\right\rangle +\left\langle RT\right\rangle -\left\langle QT\right\rangle \le 2
\end{equation}
here $\left\langle QS\right\rangle $ and similar ones are average values of  products   \textit{Q, S, R, T} and others
\begin{equation*}
\left\langle QS\right\rangle =Tr\left\{QS\rho \right\},
\end{equation*}
here \textit{$\rho $} is the density matrix of a system.  Violation of  Bell's inequality obtained for some parameters means that the system has non-local correlations, indicating the presence of entanglement.

Following to the scheme of the experiment described in \cite{Aspect}, we consider the following  situation. The ensemble of $N$ fermions looses one of indistinguishable particles so quickly that the spin has no time to be changed.  The spin state of the single fermion is analyzed by the observer A (Alice) by two devices $Q$ and $R$ (for example, Mott cells). The spin state of the rest of the composite ($N$-1)-particle  is detected by the another observer $B$ (Bob) by devices $S$ and $T$.

 Devices Q, R, S, and T are assumed to measure doubled spin projections on different axises, whose operators are
\begin{eqnarray*}
Q=\sigma _{Z}^{A}  , R=\sigma _{X}^{A} ,\\
S=2^{-1/2} (-\Theta _{Z}^{B} -\Theta _{X}^{B} ),\\
T=2^{-1/2} (\Theta _{Z}^{B} -\Theta _{X}^{B} ),
\end{eqnarray*}
here $\sigma _{Z}^{A} $, $\sigma _{X}^{A} $ are spin projection operators of the single fermion on the \textit{z} and the \textit{x} axises measured by the observer $A$, and $\Theta _{Z,X}^{B} =\sum _{i=2}^{N}\sigma _{Z,X}^{i}  $ are spin projections of the rest ($N$-1)-particle measured by the observer $B$.

After substituting operators $Q, R, S$ and $T$ and the spin density matrix $\rho ^{N} $ in the left part of the formula \eqref{15} one can obtain
\begin{eqnarray} \label{16}
{2^{1/2} \left|Tr(\sigma _{Z}^{A} \Theta _{Z}^{B} +\sigma _{X}^{A} \Theta _{X}^{B} )\rho ^{N} \right|=} \nonumber \\ {=2^{1/2} \left|Tr(\sigma _{Z}^{1} \sum _{i=2}^{N}\sigma _{Z}^{i}  +\sigma _{X}^{1} \sum _{i=2}^{N}\sigma _{X}^{i}  )\rho ^{N} \right|=} \nonumber \\ {=2^{1/2} \left|Tr\sum _{i=2}^{N}(\sigma _{Z}^{1} \sigma _{Z}^{i} ) \rho ^{N} +Tr\sum _{i=2}^{N}(\sigma _{X}^{1} \sigma _{X}^{i} ) \rho ^{N} \right|}
\end{eqnarray}

As far as products of two operators $\sigma _{Z}^{1} \sigma _{Z}^{i} $ and $\sigma _{X}^{1} \sigma _{X}^{i} $  are used only in the right part of the formula \eqref{16}, and all fermions are equivalent ones, the reduced two particle spin density matrix $\rho ^{1i} $  \eqref{9} can be used instead of the multifermion density matrix $\rho ^{N} $ \eqref{4}. Thus,
\begin{eqnarray*}
  Tr\left\{\left(\sum _{i=2}^{N}\sigma _{X}^{1} \sigma _{X}^{i}  \right)\rho ^{N} \right\}=\sum _{i=2}^{N}Tr(\sigma _{X}^{1} \sigma _{X}^{i} )\rho ^{1i} \\
  Tr\left\{\left(\sum _{i=2}^{N}\sigma _{Z}^{1} \sigma _{Z}^{i}  \right)\rho ^{N} \right\}=\sum _{i=2}^{N}Tr(\sigma _{Z}^{1} \sigma _{Z}^{i} )\rho ^{1i}
\end{eqnarray*}
and the Bell's inequality takes the form
\begin{eqnarray*}
  {\left\langle QS\right\rangle +\left\langle RS\right\rangle +\left\langle RT\right\rangle -\left\langle QT\right\rangle =} \\ {=2^{1/2} \left|\sum _{i=2}^{N}Tr\left\{\left(\sigma _{Z}^{1} \sigma _{Z}^{i} \right)\rho ^{1i} \right\}+\sum _{i=2}^{N}Tr\left\{\left(\sigma _{X}^{1} \sigma _{X}^{i} \right)\rho ^{1i} \right\}  \right|=} \\ {=2^{1/2} \left|\sum _{i=2}^{N}Tr(\sigma _{Z}^{1} \sigma _{Z}^{i} )\rho ^{1i} +\sum _{i=2}^{N}Tr(\sigma _{X}^{1} \sigma _{X}^{i} )\rho ^{1i}   \right|=2\sqrt{2} >2.}
\end{eqnarray*}
This result proves the violation of Bell's inequality for the case of the multifermion system decay. Moreover, the violation of the Bell's inequalities does not depend on the number \textit{N} of fermions, and the multifermion singlet system is similar to the two spin one.

This result could be expected from the physical point of view: the disintegration of the complex singlet particle into two fragments with spin S = 1/2 is similar to the disintegration of the two-electron system. However, the important difference between these systems  (two-spin and  multispin ones) should be noted here: in both cases the systems are entangled, but in the multifermion system  the extracted fermion spin is entangled with the whole (\textit{N}-1)-fermion system, but is  not entangled with any other fermion spin, which is left in the (\textit{N}-1)-fermion system. The considered case of the multifermion system is the example how nonentangled particles can be united in the whole entangled system.

\section{\label{sec:level1}Conclusions}

The principle of indistinguishability of particles and the Pauli's principle are proved to determine spin states of fermions uniquely, their spin correlations and entanglements of their spin states. If \textit{N}-odd fermion ensemble is in the ground state, then spin subsystems are described by density matrixes, which could be presented as sums of non-orthogonal projection operators for all possible multispin singlet states. Such presentations of spin density matrices are equivalent to the nonorthogonal decompositions of unity operators.

Multifermion systems following the Pauli's principle are shown to have entangled spin subsystems. To prove the spin entanglement in large systems the Sylvester's criterion of the matrix nonegativity  has been shown to be more convenient than the another ones as far as it does not demand  calculations of eigenvalues of large partly transposed  matrices.  Large fermion spin systems have shown can have nonentangled subsystems: for example, the 4-fermion system has partly entangled 3-spin subsystems and nonentanged 2-spin subsystems.   Due to the Pauli's principle the spin state of any two-fermion subsystem of large ensembles can be the noncoherent mixture of triplet and singlet states only. The supposed existence of pure singlet fermion subsystems are proved to be in contradiction with the Pauli's principle.

For multispin fermion ensembles the analog of the Einstein-Podolsky-Rosen experiment was analyzed in details, and the violation of the Bell's inequality was proved. Thus, the entanglement of the single fermion spin with the whole \textit{N}-1-fermion spin system was confirmed. However, in the large initial ensemble any two fermion spins were nonentangled.
\begin{acknowledgments}
Authors are thankful to Professors J. Jones (Oxford University), G. K\"{o}the (Freiburg University), G. Lesovik (Landau ITP) and S. Filippov (MIPT) for useful and helpful discussions. The financial support of Russian Foundation "Dynasty" and Orenburg University are greatly acknowledged.
\end{acknowledgments}
\bibliography{Spin}

\providecommand{\noopsort}[1]{}\providecommand{\singleletter}[1]{#1}%
\begin{thebibliography}{29}%
\makeatletter
\providecommand \@ifxundefined [1]{%
 \@ifx{#1\undefined}
}%
\providecommand \@ifnum [1]{%
 \ifnum #1\expandafter \@firstoftwo
 \else \expandafter \@secondoftwo
 \fi
}%
\providecommand \@ifx [1]{%
 \ifx #1\expandafter \@firstoftwo
 \else \expandafter \@secondoftwo
 \fi
}%
\providecommand \natexlab [1]{#1}%
\providecommand \enquote  [1]{``#1''}%
\providecommand \bibnamefont  [1]{#1}%
\providecommand \bibfnamefont [1]{#1}%
\providecommand \citenamefont [1]{#1}%
\providecommand \href@noop [0]{\@secondoftwo}%
\providecommand \href [0]{\begingroup \@sanitize@url \@href}%
\providecommand \@href[1]{\@@startlink{#1}\@@href}%
\providecommand \@@href[1]{\endgroup#1\@@endlink}%
\providecommand \@sanitize@url [0]{\catcode `\\12\catcode `\$12\catcode
  `\&12\catcode `\#12\catcode `\^12\catcode `\_12\catcode `\%12\relax}%
\providecommand \@@startlink[1]{}%
\providecommand \@@endlink[0]{}%
\providecommand \url  [0]{\begingroup\@sanitize@url \@url }%
\providecommand \@url [1]{\endgroup\@href {#1}{\urlprefix }}%
\providecommand \urlprefix  [0]{URL }%
\providecommand \Eprint [0]{\href }%
\providecommand \doibase [0]{http://dx.doi.org/}%
\providecommand \selectlanguage [0]{\@gobble}%
\providecommand \bibinfo  [0]{\@secondoftwo}%
\providecommand \bibfield  [0]{\@secondoftwo}%
\providecommand \translation [1]{[#1]}%
\providecommand \BibitemOpen [0]{}%
\providecommand \bibitemStop [0]{}%
\providecommand \bibitemNoStop [0]{.\EOS\space}%
\providecommand \EOS [0]{\spacefactor3000\relax}%
\providecommand \BibitemShut  [1]{\csname bibitem#1\endcsname}%
\let\auto@bib@innerbib\@empty
\bibitem [{\citenamefont {Schliemann}\ \emph {et~al.}(2001)\citenamefont
  {Schliemann}, \citenamefont {Cirac}, \citenamefont {Lewenstein},\ and\
  \citenamefont {Loss}}]{Schli01}%
  \BibitemOpen
  \bibfield  {author} {\bibinfo {author} {\bibfnamefont {J.}~\bibnamefont
  {Schliemann}}, \bibinfo {author} {\bibfnamefont {I.}~\bibnamefont {Cirac}},
  \bibinfo {author} {\bibfnamefont {M.}~\bibnamefont {Lewenstein}}, \ and\
  \bibinfo {author} {\bibfnamefont {D.}~\bibnamefont {Loss}},\ }\href@noop {}
  {\bibfield  {journal} {\bibinfo  {journal} {Phys. Rev. A.}\ }\textbf
  {\bibinfo {volume} {64}},\ \bibinfo {pages} {022303} (\bibinfo {year}
  {2001})}\BibitemShut {NoStop}%
\bibitem [{\citenamefont {Eckert}\ \emph {et~al.}(2002)\citenamefont {Eckert},
  \citenamefont {Schliemann}, \citenamefont {Bruss},\ and\ \citenamefont
  {Lewenstein}}]{Eckert}%
  \BibitemOpen
  \bibfield  {author} {\bibinfo {author} {\bibfnamefont {K.}~\bibnamefont
  {Eckert}}, \bibinfo {author} {\bibfnamefont {J.}~\bibnamefont {Schliemann}},
  \bibinfo {author} {\bibfnamefont {D.}~\bibnamefont {Bruss}}, \ and\ \bibinfo
  {author} {\bibfnamefont {M.}~\bibnamefont {Lewenstein}},\ }\href@noop {}
  {\bibfield  {journal} {\bibinfo  {journal} {Annals of Physics}\ }\textbf
  {\bibinfo {volume} {88}},\ \bibinfo {pages} {299} (\bibinfo {year}
  {2002})}\BibitemShut {NoStop}%
\bibitem [{\citenamefont {Amico}\ \emph {et~al.}(2008)\citenamefont {Amico},
  \citenamefont {Fazio}, \citenamefont {Osterloh},\ and\ \citenamefont
  {Vedral}}]{Amico}%
  \BibitemOpen
  \bibfield  {author} {\bibinfo {author} {\bibfnamefont {L.}~\bibnamefont
  {Amico}}, \bibinfo {author} {\bibfnamefont {L.}~\bibnamefont {Fazio}},
  \bibinfo {author} {\bibfnamefont {A.}~\bibnamefont {Osterloh}}, \ and\
  \bibinfo {author} {\bibfnamefont {V.}~\bibnamefont {Vedral}},\ }\href@noop {}
  {\bibfield  {journal} {\bibinfo  {journal} {Rev. Mod. Phys.}\ }\textbf
  {\bibinfo {volume} {80}},\ \bibinfo {pages} {517} (\bibinfo {year}
  {2008})}\BibitemShut {NoStop}%
\bibitem [{\citenamefont {Zutic}\ \emph {et~al.}(2004)\citenamefont {Zutic},
  \citenamefont {Fabian},\ and\ \citenamefont {Sarma}}]{Fabian}%
  \BibitemOpen
  \bibfield  {author} {\bibinfo {author} {\bibfnamefont {I.}~\bibnamefont
  {Zutic}}, \bibinfo {author} {\bibfnamefont {J.}~\bibnamefont {Fabian}}, \
  and\ \bibinfo {author} {\bibfnamefont {S.~D.}\ \bibnamefont {Sarma}},\
  }\href@noop {} {\bibfield  {journal} {\bibinfo  {journal} {Rev. Mod. Phys.}\
  }\textbf {\bibinfo {volume} {76}},\ \bibinfo {pages} {323} (\bibinfo {year}
  {2004})}\BibitemShut {NoStop}%
\bibitem [{\citenamefont {Valiev}\ and\ \citenamefont {Kokin}(2004)}]{Valiev}%
  \BibitemOpen
  \bibfield  {author} {\bibinfo {author} {\bibfnamefont {K.~A.}\ \bibnamefont
  {Valiev}}\ and\ \bibinfo {author} {\bibfnamefont {A.~A.}\ \bibnamefont
  {Kokin}},\ }\href@noop {} {\emph {\bibinfo {title} {Quantum Fields in Curved
  Space}}}\ (\bibinfo  {publisher} {Regular and Chaotic Dynamics},\ \bibinfo
  {year} {2004})\BibitemShut {NoStop}%
\bibitem [{\citenamefont {Kitaev}\ \emph {et~al.}(2002)\citenamefont {Kitaev},
  \citenamefont {Shen},\ and\ \citenamefont {Vayliy}}]{Kitaev}%
  \BibitemOpen
  \bibfield  {author} {\bibinfo {author} {\bibfnamefont {A.}~\bibnamefont
  {Kitaev}}, \bibinfo {author} {\bibfnamefont {A.}~\bibnamefont {Shen}}, \ and\
  \bibinfo {author} {\bibfnamefont {M.}~\bibnamefont {Vayliy}},\ }\href@noop {}
  {\emph {\bibinfo {title} {Classical and quantum computation}}}\ (\bibinfo
  {publisher} {American Mathematical Soc.},\ \bibinfo {year}
  {2002})\BibitemShut {NoStop}%
\bibitem [{\citenamefont {Cirac}\ and\ \citenamefont {Zoller}(1995)}]{Cirac95}%
  \BibitemOpen
  \bibfield  {author} {\bibinfo {author} {\bibfnamefont {J.}~\bibnamefont
  {Cirac}}\ and\ \bibinfo {author} {\bibfnamefont {P.}~\bibnamefont {Zoller}},\
  }\href@noop {} {\bibfield  {journal} {\bibinfo  {journal} {Phys. Rev. Lett.}\
  }\textbf {\bibinfo {volume} {74}},\ \bibinfo {pages} {20} (\bibinfo {year}
  {1995})}\BibitemShut {NoStop}%
\bibitem [{\citenamefont {Kilin}(1995)}]{Kilin}%
  \BibitemOpen
  \bibfield  {author} {\bibinfo {author} {\bibfnamefont {S.}~\bibnamefont
  {Kilin}},\ }\href@noop {} {\bibfield  {journal} {\bibinfo  {journal}
  {Physics-Uspekhi.}\ }\textbf {\bibinfo {volume} {169}},\ \bibinfo {pages} {5}
  (\bibinfo {year} {1995})}\BibitemShut {NoStop}%
\bibitem [{\citenamefont {Zel'dovich}\ \emph {et~al.}(1988)\citenamefont
  {Zel'dovich}, \citenamefont {Buchachenko},\ and\ \citenamefont
  {Frankevich}}]{Zel}%
  \BibitemOpen
  \bibfield  {author} {\bibinfo {author} {\bibfnamefont {Y.~B.}\ \bibnamefont
  {Zel'dovich}}, \bibinfo {author} {\bibfnamefont {A.~L.}\ \bibnamefont
  {Buchachenko}}, \ and\ \bibinfo {author} {\bibfnamefont {E.~L.}\ \bibnamefont
  {Frankevich}},\ }\href@noop {} {\bibfield  {journal} {\bibinfo  {journal}
  {Physics-Uspekhi.}\ }\textbf {\bibinfo {volume} {155}},\ \bibinfo {pages} {1}
  (\bibinfo {year} {1988})}\BibitemShut {NoStop}%
\bibitem [{\citenamefont {Bouwmeester}\ \emph {et~al.}(2000)\citenamefont
  {Bouwmeester}, \citenamefont {Ekkert},\ and\ \citenamefont
  {Zeilinger}}]{Bouw}%
  \BibitemOpen
  \bibfield  {author} {\bibinfo {author} {\bibfnamefont {D.}~\bibnamefont
  {Bouwmeester}}, \bibinfo {author} {\bibfnamefont {A.}~\bibnamefont {Ekkert}},
  \ and\ \bibinfo {author} {\bibfnamefont {A.}~\bibnamefont {Zeilinger}},\
  }\href@noop {} {\emph {\bibinfo {title} {The Physics of Quantum Information:
  Quantum Cryptography, Quantum Teleportation, Quantum Computations}}}\
  (\bibinfo  {publisher} {Springer-Verlag},\ \bibinfo {address} {Berlin},\
  \bibinfo {year} {2000})\BibitemShut {NoStop}%
\bibitem [{\citenamefont {Wang}\ and\ \citenamefont {Zanardi}(2002)}]{Wang}%
  \BibitemOpen
  \bibfield  {author} {\bibinfo {author} {\bibfnamefont {X.}~\bibnamefont
  {Wang}}\ and\ \bibinfo {author} {\bibfnamefont {P.}~\bibnamefont {Zanardi}},\
  }\href@noop {} {\bibfield  {journal} {\bibinfo  {journal} {Phys. Lett. A}\
  }\textbf {\bibinfo {volume} {301}},\ \bibinfo {pages} {1} (\bibinfo {year}
  {2002})}\BibitemShut {NoStop}%
\bibitem [{\citenamefont {Lunkes}\ \emph {et~al.}(2005)\citenamefont {Lunkes},
  \citenamefont {Brukner},\ and\ \citenamefont {Vedral}}]{Lunk05}%
  \BibitemOpen
  \bibfield  {author} {\bibinfo {author} {\bibfnamefont {C.}~\bibnamefont
  {Lunkes}}, \bibinfo {author} {\bibfnamefont {C.}~\bibnamefont {Brukner}}, \
  and\ \bibinfo {author} {\bibfnamefont {V.}~\bibnamefont {Vedral}},\
  }\href@noop {} {\bibfield  {journal} {\bibinfo  {journal} {Phys. Rev. Lett.}\
  }\textbf {\bibinfo {volume} {95}},\ \bibinfo {pages} {030503} (\bibinfo
  {year} {2005})}\BibitemShut {NoStop}%
\bibitem [{\citenamefont {Nielsen}\ and\ \citenamefont
  {Chuang}(2000)}]{Niel20}%
  \BibitemOpen
  \bibfield  {author} {\bibinfo {author} {\bibfnamefont {M.~A.}\ \bibnamefont
  {Nielsen}}\ and\ \bibinfo {author} {\bibfnamefont {I.~L.}\ \bibnamefont
  {Chuang}},\ }\href@noop {} {\emph {\bibinfo {title} {Quantum Computation and
  Information}}}\ (\bibinfo  {publisher} {Univ. Press.},\ \bibinfo {address}
  {Cambridge},\ \bibinfo {year} {2000})\BibitemShut {NoStop}%
\bibitem [{\citenamefont {Valiev}(2005)}]{Val05}%
  \BibitemOpen
  \bibfield  {author} {\bibinfo {author} {\bibfnamefont {K.~A.}\ \bibnamefont
  {Valiev}},\ }\href@noop {} {\bibfield  {journal} {\bibinfo  {journal}
  {Physics-Uspekhi}\ }\textbf {\bibinfo {volume} {175}},\ \bibinfo {pages} {1}
  (\bibinfo {year} {2005})}\BibitemShut {NoStop}%
\bibitem [{\citenamefont {Vedral}(2003)}]{Ved03}%
  \BibitemOpen
  \bibfield  {author} {\bibinfo {author} {\bibfnamefont {V.}~\bibnamefont
  {Vedral}},\ }\href@noop {} {\bibfield  {journal} {\bibinfo  {journal}
  {Central Eur. J. Phys.}\ }\textbf {\bibinfo {volume} {1}} (\bibinfo {year}
  {2003})}\BibitemShut {NoStop}%
\bibitem [{\citenamefont {Oh}\ and\ \citenamefont {Kim}(2004)}]{Oh04}%
  \BibitemOpen
  \bibfield  {author} {\bibinfo {author} {\bibfnamefont {S.}~\bibnamefont
  {Oh}}\ and\ \bibinfo {author} {\bibfnamefont {J.}~\bibnamefont {Kim}},\
  }\href@noop {} {\bibfield  {journal} {\bibinfo  {journal} {Phys. Rev. A}\
  }\textbf {\bibinfo {volume} {69}},\ \bibinfo {pages} {054305} (\bibinfo
  {year} {2004})}\BibitemShut {NoStop}%
\bibitem [{\citenamefont {Landau}\ and\ \citenamefont
  {Lifshitz}(1974)}]{Landau}%
  \BibitemOpen
  \bibfield  {author} {\bibinfo {author} {\bibfnamefont {L.~D.}\ \bibnamefont
  {Landau}}\ and\ \bibinfo {author} {\bibfnamefont {E.~M.}\ \bibnamefont
  {Lifshitz}},\ }\href@noop {} {\emph {\bibinfo {title} {Quantum mechanics}}}\
  (\bibinfo  {publisher} {Nauka},\ \bibinfo {address} {Moscow},\ \bibinfo
  {year} {1974})\BibitemShut {NoStop}%
\bibitem [{\citenamefont {Blum}()}]{Blum}%
  \BibitemOpen
  \bibfield  {author} {\bibinfo {author} {\bibfnamefont {K.}~\bibnamefont
  {Blum}},\ }\href@noop {} {\emph {\bibinfo {title} {Density Matrix Theory and
  Applications}}}\ (\bibinfo  {publisher} {Mir},\ \bibinfo {address}
  {Moscow})\BibitemShut {NoStop}%
\bibitem [{\citenamefont {Rumer}\ and\ \citenamefont {Fet}(1970)}]{Rumer}%
  \BibitemOpen
  \bibfield  {author} {\bibinfo {author} {\bibfnamefont {Y.~B.}\ \bibnamefont
  {Rumer}}\ and\ \bibinfo {author} {\bibfnamefont {.~I.}\ \bibnamefont {Fet}},\
  }\href@noop {} {\emph {\bibinfo {title} {The theory of unitary symmetry}}}\
  (\bibinfo  {publisher} {Nauka},\ \bibinfo {address} {Moscow},\ \bibinfo
  {year} {1970})\BibitemShut {NoStop}%
\bibitem [{\citenamefont {Aldoshin}\ \emph {et~al.}(2008)\citenamefont
  {Aldoshin}, \citenamefont {Feldman},\ and\ \citenamefont {Yurishev}}]{Aldos}%
  \BibitemOpen
  \bibfield  {author} {\bibinfo {author} {\bibfnamefont {S.~M.}\ \bibnamefont
  {Aldoshin}}, \bibinfo {author} {\bibfnamefont {E.}~\bibnamefont {Feldman}}, \
  and\ \bibinfo {author} {\bibfnamefont {M.~A.}\ \bibnamefont {Yurishev}},\
  }\href@noop {} {\bibfield  {journal} {\bibinfo  {journal} {JETP}\ }\textbf
  {\bibinfo {volume} {134}},\ \bibinfo {pages} {5} (\bibinfo {year}
  {2008})}\BibitemShut {NoStop}%
\bibitem [{\citenamefont {Peres}(1996)}]{Peres}%
  \BibitemOpen
  \bibfield  {author} {\bibinfo {author} {\bibfnamefont {A.}~\bibnamefont
  {Peres}},\ }\href@noop {} {\bibfield  {journal} {\bibinfo  {journal} {Phys.
  Rev. Lett.}\ }\textbf {\bibinfo {volume} {77}},\ \bibinfo {pages} {1413}
  (\bibinfo {year} {1996})}\BibitemShut {NoStop}%
\bibitem [{\citenamefont {Horodecki}\ \emph {et~al.}(1996)\citenamefont
  {Horodecki}, \citenamefont {Horodecki},\ and\ \citenamefont
  {Horodecki}}]{Horod96}%
  \BibitemOpen
  \bibfield  {author} {\bibinfo {author} {\bibfnamefont {M.}~\bibnamefont
  {Horodecki}}, \bibinfo {author} {\bibfnamefont {P.}~\bibnamefont
  {Horodecki}}, \ and\ \bibinfo {author} {\bibfnamefont {R.}~\bibnamefont
  {Horodecki}},\ }\href@noop {} {\bibfield  {journal} {\bibinfo  {journal}
  {Phys. Lett. A}\ }\textbf {\bibinfo {volume} {223}},\ \bibinfo {pages} {1}
  (\bibinfo {year} {1996})}\BibitemShut {NoStop}%
\bibitem [{\citenamefont {Vidal}\ and\ \citenamefont {Werner}(2002)}]{Vidal02}%
  \BibitemOpen
  \bibfield  {author} {\bibinfo {author} {\bibfnamefont {G.}~\bibnamefont
  {Vidal}}\ and\ \bibinfo {author} {\bibfnamefont {R.~F.~A.}\ \bibnamefont
  {Werner}},\ }\href@noop {} {\bibfield  {journal} {\bibinfo  {journal} {Phys.
  Rev. A}\ }\textbf {\bibinfo {volume} {65}},\ \bibinfo {pages} {032314}
  (\bibinfo {year} {2002})}\BibitemShut {NoStop}%
\bibitem [{\citenamefont {Nielsen}(1999)}]{Niel99}%
  \BibitemOpen
  \bibfield  {author} {\bibinfo {author} {\bibfnamefont {M.~A.}\ \bibnamefont
  {Nielsen}},\ }\href@noop {} {\bibfield  {journal} {\bibinfo  {journal} {Phys.
  Rev. Lett.}\ }\textbf {\bibinfo {volume} {83}},\ \bibinfo {pages} {436}
  (\bibinfo {year} {1999})}\BibitemShut {NoStop}%
\bibitem [{\citenamefont {Belousov}\ and\ \citenamefont
  {Manko}(2004)}]{Belous}%
  \BibitemOpen
  \bibfield  {author} {\bibinfo {author} {\bibfnamefont {Y.~M.}\ \bibnamefont
  {Belousov}}\ and\ \bibinfo {author} {\bibfnamefont {V.~I.}\ \bibnamefont
  {Manko}},\ }\href@noop {} {\emph {\bibinfo {title} {Density Matrix.
  Presentation and application in statistical mechanics}}}\ (\bibinfo
  {publisher} {MPTI},\ \bibinfo {address} {Moscow},\ \bibinfo {year}
  {2004})\BibitemShut {NoStop}%
\bibitem [{\citenamefont {Gantmaher}(1966)}]{Gant}%
  \BibitemOpen
  \bibfield  {author} {\bibinfo {author} {\bibfnamefont {F.~R.}\ \bibnamefont
  {Gantmaher}},\ }\href@noop {} {\emph {\bibinfo {title} {Matrix theory}}}\
  (\bibinfo  {publisher} {Nauka},\ \bibinfo {address} {Moscow},\ \bibinfo
  {year} {1966})\BibitemShut {NoStop}%
\bibitem [{\citenamefont {Einstein}\ \emph {et~al.}(1935)\citenamefont
  {Einstein}, \citenamefont {Podolsky},\ and\ \citenamefont {Rosen}}]{epr}%
  \BibitemOpen
  \bibfield  {author} {\bibinfo {author} {\bibfnamefont {A.}~\bibnamefont
  {Einstein}}, \bibinfo {author} {\bibfnamefont {{\relax Yu}.}~\bibnamefont
  {Podolsky}}, \ and\ \bibinfo {author} {\bibfnamefont {N.}~\bibnamefont
  {Rosen}} (\bibinfo {collaboration} {EPR}),\ }\href@noop {} {\bibfield
  {journal} {\bibinfo  {journal} {Phys.\ Rev.}\ }\textbf {\bibinfo {volume}
  {47}},\ \bibinfo {pages} {777} (\bibinfo {year} {1935})}\BibitemShut
  {NoStop}%
\bibitem [{\citenamefont {Clauser}\ \emph {et~al.}(1969)\citenamefont
  {Clauser}, \citenamefont {Horne}, \citenamefont {Shimony},\ and\
  \citenamefont {Holt}}]{Claus}%
  \BibitemOpen
  \bibfield  {author} {\bibinfo {author} {\bibfnamefont {J.~F.}\ \bibnamefont
  {Clauser}}, \bibinfo {author} {\bibfnamefont {M.~A.}\ \bibnamefont {Horne}},
  \bibinfo {author} {\bibfnamefont {A.}~\bibnamefont {Shimony}}, \ and\
  \bibinfo {author} {\bibfnamefont {R.~A.}\ \bibnamefont {Holt}},\ }\href@noop
  {} {\bibfield  {journal} {\bibinfo  {journal} {Phys. Rev. Lett.}\ }\textbf
  {\bibinfo {volume} {23}},\ \bibinfo {pages} {880} (\bibinfo {year}
  {1969})}\BibitemShut {NoStop}%
\bibitem [{\citenamefont {Aspect}\ \emph {et~al.}(1982)\citenamefont {Aspect},
  \citenamefont {Grangier},\ and\ \citenamefont {Roger}}]{Aspect}%
  \BibitemOpen
  \bibfield  {author} {\bibinfo {author} {\bibfnamefont {A.}~\bibnamefont
  {Aspect}}, \bibinfo {author} {\bibfnamefont {P.}~\bibnamefont {Grangier}}, \
  and\ \bibinfo {author} {\bibfnamefont {G.}~\bibnamefont {Roger}},\
  }\href@noop {} {\bibfield  {journal} {\bibinfo  {journal} {Phys. Rev. Lett.}\
  }\textbf {\bibinfo {volume} {49}},\ \bibinfo {pages} {91} (\bibinfo {year}
  {1982})}\BibitemShut {NoStop}%
\end{thebibliography}%

\end{document}